\documentclass[final]{aipproc}

\layoutstyle{6x9}

\begin{document}

\title{GRB 050315: A step in the proof of the uniqueness of the overall GRB structure}


\author{R. Ruffini}{
  address={ICRANet and ICRA, Piazzale della Repubblica 10, I-65100 Pescara, Italy.},
  altaddress={Dipartimento di Fisica, Universit\`a ``La Sapienza'', Piazzale Aldo Moro 5, I-00185 Roma, Italy.}
}

\author{M.G. Bernardini}{
  address={ICRANet and ICRA, Piazzale della Repubblica 10, I-65100 Pescara, Italy.},
  altaddress={Dipartimento di Fisica, Universit\`a ``La Sapienza'', Piazzale Aldo Moro 5, I-00185 Roma, Italy.}
}

\author{C.L. Bianco}{
  address={ICRANet and ICRA, Piazzale della Repubblica 10, I-65100 Pescara, Italy.},
  altaddress={Dipartimento di Fisica, Universit\`a ``La Sapienza'', Piazzale Aldo Moro 5, I-00185 Roma, Italy.}
}

\author{P. Chardonnet}{
  address={ICRANet and ICRA, Piazzale della Repubblica 10, I-65100 Pescara, Italy.},
  altaddress={Universit\'e de Savoie, LAPTH - LAPP, BP 110, F-74941 Annecy-le-Vieux Cedex, France.}
}

\author{F. Fraschetti}{
  address={ICRANet and ICRA, Piazzale della Repubblica 10, I-65100 Pescara, Italy.},
  altaddress={Osservatorio Astronomico di Brera, via Bianchi 46, I-23807 Merate (LC), Italy.}
}

\author{R. Guida}{
  address={ICRANet and ICRA, Piazzale della Repubblica 10, I-65100 Pescara, Italy.},
  altaddress={Dipartimento di Fisica, Universit\`a ``La Sapienza'', Piazzale Aldo Moro 5, I-00185 Roma, Italy.}
}

\author{S.-S. Xue}{
  address={ICRANet and ICRA, Piazzale della Repubblica 10, I-65100 Pescara, Italy.},
  altaddress={Dipartimento di Fisica, Universit\`a ``La Sapienza'', Piazzale Aldo Moro 5, I-00185 Roma, Italy.}
}

\begin{abstract}
Using the Swift data of GRB 050315, we progress in proving the uniqueness of our theoretically predicted Gamma-Ray Burst (GRB) structure as composed by a proper-GRB, emitted at the transparency of an electron-positron plasma with suitable baryon loading, and an afterglow comprising the ``prompt radiation'' as due to external shocks. Detailed light curves for selected energy bands are theoretically fitted in the entire temporal region of the Swift observations ranging over $10^6$ seconds.
\end{abstract}

\maketitle

\section{Introduction}

GRB 050315 \citep{va05} has been triggered and located by the BAT instrument \citep{b04,ba05} on board of the {\em Swift} satellite \citep{ga04} at 2005-March-15 20:59:42 UT \citep{pa05}. The narrow field instrument XRT \citep{bua04,bua05} began observations $\sim 80$ s after the BAT trigger, one of the earliest XRT observations yet made, and continued to detect the source for $\sim 10$ days \citep{va05}. The spectroscopic redshift has been found to be $z = 1.949$ \citep{kb05}. We present here the first results of the fit of this source in the framework of our theoretical model and point out the new step toward the uniqueness of the explanation of the overall GRB structure made possible by the Swift data of this source.

\section{Our theoretical model}

GRB 050315 observations find a direct explanation in our theoretical model \citep[see][and references therein]{rlet1,rlet2,rubr,rubr2,EQTS_ApJL2,PowerLaws}. We determine the values of the two free parameters which characterize our model: the total energy stored in the Dyadosphere $E_{dya}$ and the mass of the baryons left by the collapse $M_Bc^2 \equiv B E_{dya}$. We follow the expansion of the pulse, composed by the electron-positron plasma initially created by the vacuum polarization process in the Dyadosphere. The plasma self-propels itself outward and engulfs the baryonic remnant left over by the collapse of the progenitor star. As such pulse reaches transparency, the Proper Gamma-Ray Burst (P-GRB) is emitted \citep{rswx99,rswx00,rlet2}. The remaining accelerated baryons, interacting with the interstellar medium (ISM), produce the afterglow emission. The ISM is described by the two additional parameters of the theory: the average particle number density $<n_{ISM}>$ and the ratio $<\mathcal{R}>$ between the effective emitting area and the total area of the pulse \citep{spectr1}, which take into account the ISM filamentary structure \citep{fil}.

The luminosity in fixed energy bands is evaluated integrating over the equitemporal surfaces \citep[EQTSs, see][]{EQTS_ApJL,EQTS_ApJL2}, computed using the exact solutions of the afterglow equations of motion \citep{PowerLaws}, the energy density released due to the totally inelastic collisions of the accelerated baryons with the ISM measured in the co-moving frame, duly boosted in the observer frame. In the reference frame co-moving with the accelerated baryonic matter, the radiation produced by this interaction of the ISM with the front of the expanding baryonic shell is assumed to have a thermal spectrum \citep{spectr1}.

We reproduce correctly in several GRBs and in this specific case (see e.g. Figs. \ref{15-350}--\ref{global}) the observed time variability of the prompt emission as well as the remaining part of the afterglow \citep[see e.g.][and references therein]{r02,rubr,rubr2,031203}. The radiation produced by the interaction of the accelerated baryons with the ISM agrees with observations both for intensity and time structure.

As shown in previous cases (GRB 991216 \citep{rubr,beam}, GRB 980425 \citep{cospar02}, GRB 030329 \citep{mg10grazia}, GRB 031203 \citep{031203}), also for GRB 050315, using the correct equations of motion, there is no need to introduce a collimated emission to fit the afterglow observations.

The major difference between our theoretical model and the ones in the current literature \citep[see e.g.][and references therein]{p04} is that what is usually called ``prompt emission'' in our case coincides with the peak of the afterglow emission and is not due to a different physical process \citep{rlet2}. The verification of this prediction has been up to now tested in a variety of sources like GRB 991216 \citep{rubr}, GRB 980425 \citep{cospar02}, GRB 030329 \citep{mg10grazia}, GRB 031203 \citep{031203}. However, in all such sources the observational data were available only during the prompt emission and the latest afterglow phases, leaving all the in-between evolution undetermined. Now, thanks to the superb data provided by the Swift satellite, we are finally able to confirm, by direct confrontation with the observational data, our theoretical predictions on the GRB structure \citep{rlet2} with a detailed fit of the complete afterglow light curve of GRB 050315, from the peak (i.e. from the so-called ``prompt emission'') all the way to the latest phases without any gap in the observational data.

\section{GRB 991216}

A basic feature of our model consists in a sharp distinction between two different components in the GRB structure: the proper GRB (P-GRB), emitted at the moment of transparency, followed by an afterglow completely described by external shocks and composed of three different regimes. The first afterglow regime corresponds to a bolometric luminosity monotonically increasing with the photon detector arrival time, corresponding to a substantially constant Lorentz gamma factor of the accelerated baryons. The second regime consists of the bolometric luminosity peak, corresponding to the ``knee'' in the decreasing phase of the baryonic Lorentz gamma factor. The third regime corresponds to a bolometric luminosity decreasing with arrival time, corresponding to the late deceleration of the Lorentz gamma factor.

\begin{figure}
\includegraphics[width=0.9\hsize,clip]{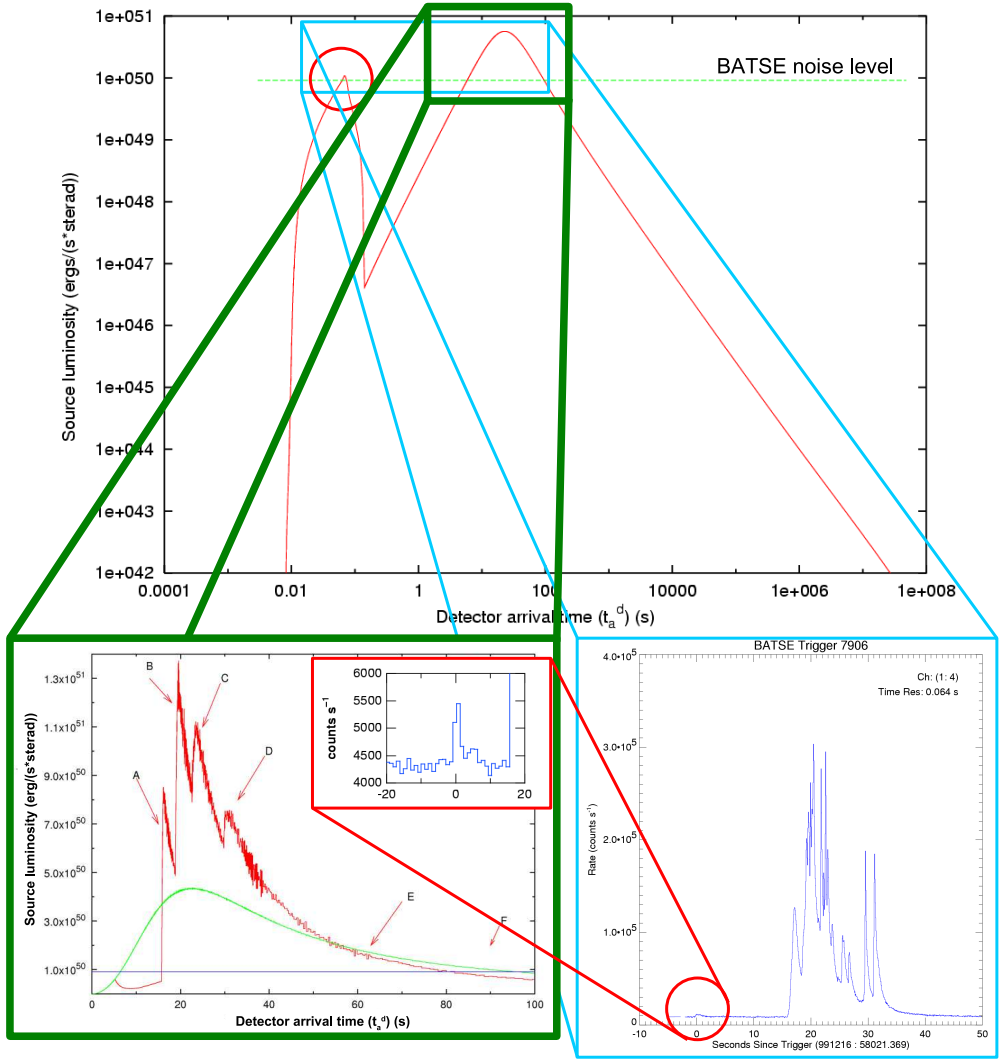}
\caption{This picture shows our prediction on the GRB structure based on the analysis of GRB 991216 \citep{rlet2}. In the main panel there is the bolometic light curve computed using our model and composed by the P-GRB and the afterglow, together with the BATSE noise level. In the lower right panel there is represented the BATSE observation of the prompt emission \citep[see][]{grblc99,brbr99}, with the clear identification of the observed ``main burst'' with the peak of the theoretical afterglow light curve and of the observed ``precursor'' with the theoretically predicted P-GRB (see enlargement). In the lower left panel is represented our theoretical fit of the BATSE observations of the afterglow peak using an inhomogeneous ISM. Details in \citet{r02,rubr,rubr2}.}
\label{991216}
\end{figure}

In some sources the P-GRB is under the observability threshold. In \citet{rlet2} we have chosen as a prototype the source GRB 991216 which clearly shows the existence of this two components. Both the relative intensity of the P-GRB to the peak of the afterglow, as well as their corresponding temporal lag, have been theoretically predicted within a few percent (see Fig. 11 in \citet{rubr}). The continuous line in the main panel of Fig. \ref{991216} corresponds to a constant ISM density averaged over the entire afterglow. The structured curve, shown in the bottom left panel, corresponds to ISM density inhomogeneities which are assumed for simplicity to be spherically symmetric \citep{r02}. Clearly, a more precise description of the BATSE light curve (e.g. the two sharp spikes at $\sim 30$ s) will need a more refined 3-dimensional description of the ISM filamentary structure \citep{fil}.

This same approximation of spherically symmetric description of the ISM inhomogeneities is in the following adopted for GRB 050315, and is sufficient to clearly outline the general behavior of the luminosity vs. photon detector arrival time in selected energy bands.

\section{The fit of the observations}

\begin{figure}
\includegraphics[width=0.9\hsize,clip]{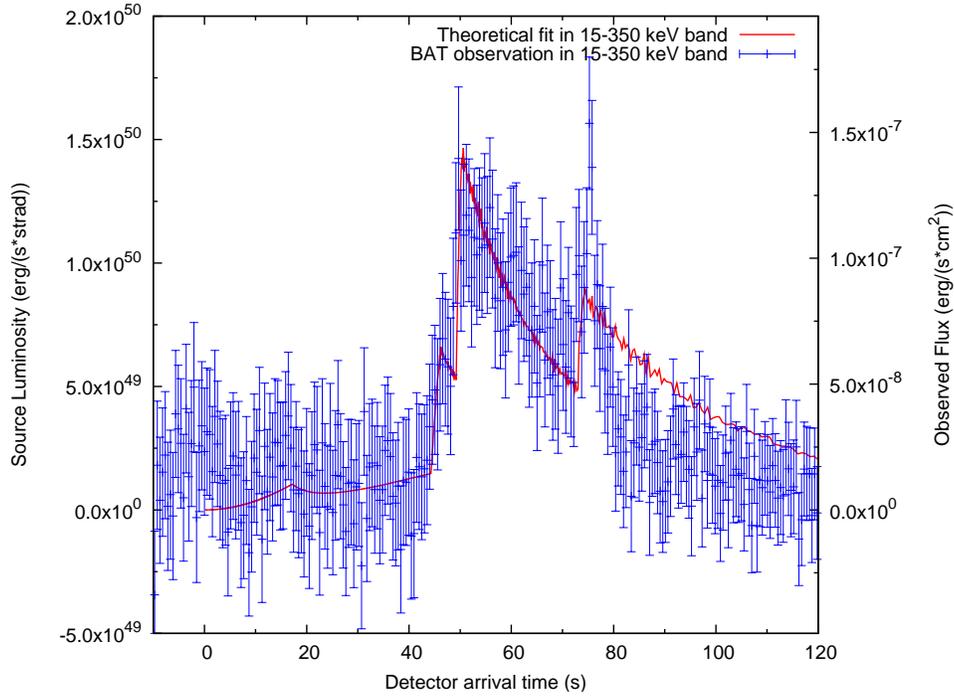}
\caption{Theoretical fit (red line), computed using our model, of the BAT observations (blue points) of GRB 050315 in the $15$--$350$ keV energy band \citep{va05}.}
\label{15-350}
\end{figure}

\begin{figure}
\includegraphics[width=0.9\hsize,clip]{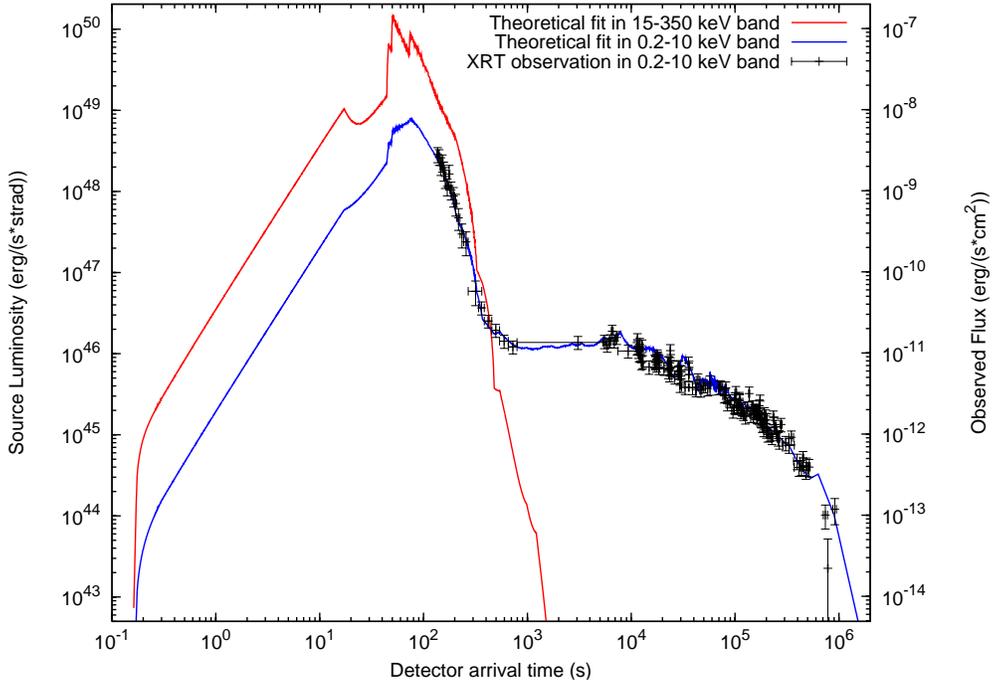}
\caption{Theoretical fit (blue line), computed using our model, of the XRT observations (black points) of GRB 050315 in the $0.2$--$10$ keV energy band \citep{va05}. The theoretical fit of the BAT observations (see Fig. \ref{15-350}) in the $15$--$350$ keV energy band is also represented (red line).}
\label{global}
\end{figure}

The best fit of the observational data leads to a total energy of the Dyadosphere $E_{dya} = 1.47\times 10^{53} erg$  \citep[the observational Swift $E_{iso}$ is $> 2.62\times 10^{52}$ erg, see Ref. ][]{va05}, so that the plasma is created between the radii $r_1 = 5.88\times 10^6$ cm and $r_2 = 1.74 \times 10^8$ cm with an initial temperature $T = 2.05 MeV$ and a total number of pairs $N_{e^+e^-} = 7.93\times 10^{57}$. The amount of baryonic matter in the remnant is assumed to be such that $B = 4.55 \times 10^{-3}$. The transparency point and the P-GRB emission occurs then with an initial Lorentz gamma factor of the accelerated baryons $\gamma_\circ = 217.81$ and at a distance $r = 1.32 \times 10^{14}$ cm from the Black Hole. The interstellar medium (ISM) parameters that we assume to best fit the observational data are: $<n_{ism}>= 0.121$ particles/cm$^{3}$ and $<R> = 2.05 \times 10^{-6}$. The ISM density contrast is found to be $\Delta \rho / \rho \sim 10^2$ on a scale of $5.0 \times 10^{16}$ cm.

In Figs. \ref{15-350} and \ref{global} we represent the theoretically computed GRB 050315 light curves, respectively in the $15$--$350$ keV and in the $0.2$-$10$ keV energy bands, which we obtained using our model, together with the corresponding data observed respectively by the BAT and the XRT instruments on board of the {\em Swift} satellite \citep{va05}. For completeness, in Fig. \ref{global} is also represented the theoretically computed $15$--$350$ keV light curve of Fig. \ref{15-350}, but not the BAT observational data to not overwhelm the picture too much.

The very good agreement between the theoretical curves and the observations is a most stringent proof of our predictions on the GRB structure \citep{rlet2}.

It goes without saying that also in the case of GRB 050315 a more detailed correspondence between the theory and the temporal fine structure of the BAT observational light curve could be achieved with a full 3-dimensional description of the ISM filamentary structure \citep{fil}.

\section{Conclusions}

In view of the above results, which clearly fits the overall luminosity in fixed energy bands, we will return in a forthcoming publication to the identification of the P-GRB using our theoretically predicted values for both its intensity and its time lag relative to the afterglow peak. We will also address the spectral analysis which is a most powerful theoretical prediction in order to evidence the continuity between the ``prompt radiation'' and the late phases of the afterglow and so to prove the uniqueness of the overall GRB structure. 
 
\begin{theacknowledgments}
We thank P. Banat, G. Chincarini, A. Moretti and S. Vaughan for their help in the analysis of the observational data.
\end{theacknowledgments}

\end{document}